\documentclass[%
prresearch,
reprint,
superscriptaddress,
 amsmath,amssymb,
 aps,
floatfix,
longbibliography,
]{revtex4-2}
\usepackage{cprotect}
\usepackage{bm}
\usepackage{graphicx,xcolor}
\usepackage{dcolumn}
\usepackage{bm}
\usepackage{physics}
\def\vec#1{\boldsymbol #1}
\usepackage{color}

\newcommand{\isspaffiliation}{The Institute for Solid State Physics,~The University of Tokyo, Kashiwa, Chiba 277-8581, Japan}

\begin{document}

\preprint{APS/123-QED}

\title{Combined x-ray diffraction, electrical resistivity, and $ab$ $initio$ study of (TMTTF)$_2$PF$_6$ \\ under pressure: implications to the unified phase diagram}


\author{Miho Itoi}
\thanks{The authors contributed to this work equally.}
\affiliation{Department of Natural Sciences, Faculty of Science and Engineering, Tokyo City University, Setagaya, Tokyo, 158-8557, Japan}
\author{Kazuyoshi Yoshimi}
\thanks{The authors contributed to this work equally.}
\affiliation{\isspaffiliation}
\author{Haming Ma}
\affiliation{\isspaffiliation}
\author{Takahiro Misawa}
\affiliation{\isspaffiliation}
\affiliation{Beijing Academy of Quantum Information Sciences, Haidian District, Beijing 100193, China}
\author{Takao Tsumuraya}
\affiliation{Magnesium Research Center, Kumamoto University, Kumamoto, Kumamoto 860-8555, Japan}
\author{Dilip Bhoi}
\affiliation{\isspaffiliation}
\author{Tokutaro Komatsu}
\affiliation{Division of Natural Sciences, Nihon University School of Medicine, Itabashi, Tokyo 173-8610, Japan}
\author{Hatsumi Mori}
\affiliation{\isspaffiliation}
\author{Yoshiya Uwatoko}
\affiliation{\isspaffiliation}
\author{Hitoshi Seo}
\affiliation{Condensed Matter Theory Laboratory, RIKEN, Wako, Saitama 351-0198, Japan}
\affiliation{Center for Emergent Matter Science (CEMS), RIKEN, Wako, Saitama 351-0198, Japan}

\date{\today}

\begin{abstract}
We present a combined experimental and theoretical study on the quasi-one-dimensional organic conductor (TMTTF)$_2$PF$_6$, and elucidate the variation of its physical properties under pressure.
We fully resolve the crystal structure by single crystal x-ray diffraction measurements using a diamond anvil cell up to 8~GPa, and based on the structural data, we perform first-principles density-functional theory calculations and derive the $ab$ $initio$ extended Hubbard-type Hamiltonians. 
Furthermore, we compare the behavior of the resistivity measured up to 3~GPa using a BeCu clamp-type cell and the ground state properties of the obtained model numerically calculated by the many-variable variational Monte Carlo method. 
Our main findings are as follows: i) The crystal was rapidly compressed up to about 3~GPa where the volume drops to 80\% and gradually varies down to 70\% at 8~GPa.  
The transfer integrals increase following such behavior whereas the screened Coulomb interactions decrease, resulting in a drastic reduction of correlation effect. 
ii) The degree of dimerization in the intrachain transfer integrals, as the result of the decrease in structural dimerization together with the change in the intermolecular configuration, almost disappears above 4~GPa; the interchain transfer integrals also show characteristic variations under pressure. 
iii) The results of identifying the characteristic temperatures in the resistivity and the charge and spin orderings in the calculations show an overall agreement: The charge ordering sensitively becomes unstable above 1~GPa, while the spin ordering survives up to higher pressures.
These results shed light on the similarities and differences between applying external pressure and substituting the chemical species (chemical pressure).
\end{abstract}

\maketitle

\section{Introduction}
Organic conductors, whose electric conductivity is carried by the conduction bands originated from the molecular orbital of the composing molecules, have been studied as a prototypical playground of electrons interacting in low-dimensional systems~\cite{Ishiguro_book, Lebed_book}. 
Their versatility in physical phenomena and phase transitions into different symmetry-broken states can be tuned systematically, taking advantage of their capability of chemically substituting the constituent molecules or counterions, and also of their structural softness resulting in relatively large deformation by applying external pressure~\cite{Jerome_ChemRev, Miyagawa_ChemRev, Brown_review}. 
These ways of controlling the phases are called chemical and physical pressure, respectively. 
As for the latter strategy, aiming at continuously varying the governing factors of the physical properties, techniques using different kinds of apparatus have been applied~\cite{Adachi_doi:10.1021/ja0001254, DJaccard_2001, Taniguchi_doi:10.1143/JPSJ.72.468, ITOI2010S594, Taniguchi_Review}. 

Once the variation of physical properties among certain classes of materials was recognized to be continuous in many cases, unified phase diagrams have been proposed to summarize them by connecting pressure-temperature phase diagrams for different salts. 
There, the chemical and the physical pressures are combined as a common axis, 
and the sequences of the observed phenomena and emergent phases are put together. 
Such integration is pursued based on the assumption that the two approaches lead to similar effects, naively, to tune the volume of the unit cell resulting in the variation of microscopic parameters while keeping the important ingredients. 
Here, we propose a direct way to investigate this issue without ambiguity, by combining precise experiments and an \textit{ab initio} calculation scheme,  which can provide feedback to the design of materials exhibiting novel properties through qualitative evaluation of a system under pressure. 

We focus on an early well-established example, i.e., the unified phase diagram  of the family composed of TMTTF or TMTSF molecules~\cite{BECHGAARD19801119, Jerome_science, Brown_review}. 
These molecules form a 2:1 composition with different counter anions, {\it X}: they are written as (TM)$_2${\it X}, TM standing for TMTTF/TMTSF.
In the low-pressure side of the phase diagram, strongly correlated physics is displayed. 
Owing to the presence of structural dimerization along the TM chain axis in the 3/4-filled conduction band leading to an effective 1/2-filling situation, a dimer-type Mott (DM) insulating state is realized by the on-site Coulomb repulsion ($U$) within the molecular orbital~\cite{Emery_PRL1982}. 
Furthermore, the intersite Coulomb repulsion brings about instability toward charge order (CO)~\cite{seo_JPSJreview, Chow_2000, CO-PhysRevLett.86.4080, Nad2006, Yoshimi_2012}. 
By increasing pressure, the system turns metallic, and a phase competition between spin density wave (SDW) and anisotropic superconductivity is manifested~\cite{Takigawa_1987, Hasegawa_1987, Lee_PhysRevB.62.R14669, usr_PhysRevLett.110.107005}. 
Here the chemical pressure is interpreted in two ways~\cite{yoshimi2022}: 
i) TMTTF salts are more correlated than TMTSF salts since the spacial extension of molecular orbital in the former is smaller, leading to smaller transfer integrals and therefore resulting in effectively larger correlation.
ii) Smaller counteranion {\it X} makes the packing of TM molecules closer  resulting in larger transfer integrals and then in a wider bandwidth. 

On the other hand, to conduct measurements under applied pressure, different pressure-generating techniques have been applied, depending on the purpose of the experiment.
To fine-tune the pressure in order to investigate the details of 
phase competitions and coexistences within high-precision He gas pressure technique has been applied~\cite{Kagawa_PRB2004, ITOI2010S594} but is limited to the small range of pressure.
A contrasting motivation is to apply larger pressure and see, for example, whether the sequence of the phase diagram is reproduced from the low-pressure side up to the high-pressure regime such as the superconducting phase and beyond~\cite{Adachi_doi:10.1021/ja0001254, DJaccard_2001, kano2012anisotropy}. 

One problem when interpreting the results and comparing them with the unified phase diagram is the lack of structural data under pressure. 
Since many of the physical properties have been discussed based on microscopic models, mostly using Hubbard-type models, it would be beneficial to evaluate the model parameters and discuss the predictions from theory in comparison with the experimental results~\cite{Yoshioka_crystals_review}. 
A few attempts to extract such parameters under pressure have been made, for example, using the cell parameters under pressure and optimizing the internal atomic positions by first-principles calculations~\cite{Jacko2013, Rose_2013, pashkin2009pressure}. 
Nevertheless, it is known that structural optimization for molecular materials has problems itself, such as the choice of van der Waals potentials. 

In this work, to overcome these problems, we fully resolve the crystal structure under a wide range of pressure (0 $\leq P \leq $ 8~GPa). 
Then, using the data, we apply an {\it ab initio} model calculation scheme: for each pressure, we derive the parameters for the extended Hubbard-type model, i.e., the intermolecular transfer integrals and the Coulomb interactions from first-principles band calculations, and numerically determine the ground state of the model using a highly-accurate solver. 
By comparing the theoretical results and the experimentally measured temperature dependence of electrical resistivity, we discuss the implications of the unified phase diagram of TM$_2${\it X}, especially the similarities and differences between the chemical and physical pressure effects. 

\section{Experiments}\label{experiments}
\subsection{Material and Experimental methods}
Single crystals of (TMTTF)$_2$PF$_6$ were obtained by a conventional electrochemical technique~\cite{BECHGAARD19801119}. 
The resistivity was measured along the {\it a} axis by the 4-probe method, using a BeCu clamp-type pressure cell up to 3~GPa.
Our aim here is to measure the detailed behavior of resistivity at the pressure region where the variation of phase competitions and co-existences are argued to take place sensitively, rather than to reach the superconducting phase at around 4 -- 5~GPa~\cite{Adachi_doi:10.1021/ja0001254, DJaccard_2001, araki2007electrical, kano2012anisotropy}.
The applied pressure was determined by the superconducting temperature of a lead wire, which was placed in the BeCu cell with the sample. 
The resistance data were collected at interval of every 0.3 $\sim$ 0.4~GPa by PPMS (Quantum design) with the sequence procedure of cooling the sample down first and then heating it up with 0.4~K/min temperature control rate, to avoid the sample cracking. 

On the other hand, single crystal x-ray diffraction (XRD) measurement under high pressure was performed with XtaLAB (Rigaku) AFC11 (RCD3): quarter-chi single diffractometer.
A diamond anvil cell (DAC) was used as the pressure generator, and the sample was placed in a Re gasket hole of $\phi$ 300 $\mu \rm{m} \times 80 ~\mu \rm{m}$, in which the pressure-transmitting medium of Methanol and Ethanol mixture of 1:1 was filled. The sample size used for x-ray measurement under pressure was 0.16 × 0.14 × 0.07 mm$^3$.
The applied pressure was calibrated by Ruby R$_1$ fluorescence line~\cite{yotaru_10.1063/1.325277}. 

The x-ray radiation wavelength was 0.71073 \r{A} (Mo K$\alpha$).
The data were measured using $\omega$ scans of 0.5$^\circ$ per frame for 1.0 s, and all data were collected at room temperature. 
The XRD patterns were indexed and the total number of runs and images were based on the strategy calculation from the program CrysAlisPro (Rigaku, V1.171.41.122a, 2021).
The maximum resolution that was achieved was $\theta_{max} \sim  29^{\circ}$.   
The crystal data was solved with the ShelXT (Sheldrick, 2015) structure solution program using the Intrinsic Phasing solution method and by using Olex2 (Dolomanov et al., 2009) as the graphical interface. 
The model was refined with version 2018/3 of ShelXL 2018/3 (Sheldrick, 2015) using Least Squares minimization. 
When refining the crystal structure under pressure, all bond lengths and angles between atoms in the TMTTF molecule and PF$_6$ were not constrained in the analysis. However, because the use of DAC would result in insufficient number of reflections to perform a complete structural analysis, the analysis of the anisotropic thermal factors of the TMTTF molecule and PF$_6$ anion was avoided, and all atoms were restricted to isothermal factors.
The applied pressures during the XRD measurements and other experimental conditions under several pressures are presented in Appendix A. 

\subsection{Electrical resistivity}

\begin{figure}
\begin{center} 
\includegraphics[width=85mm]{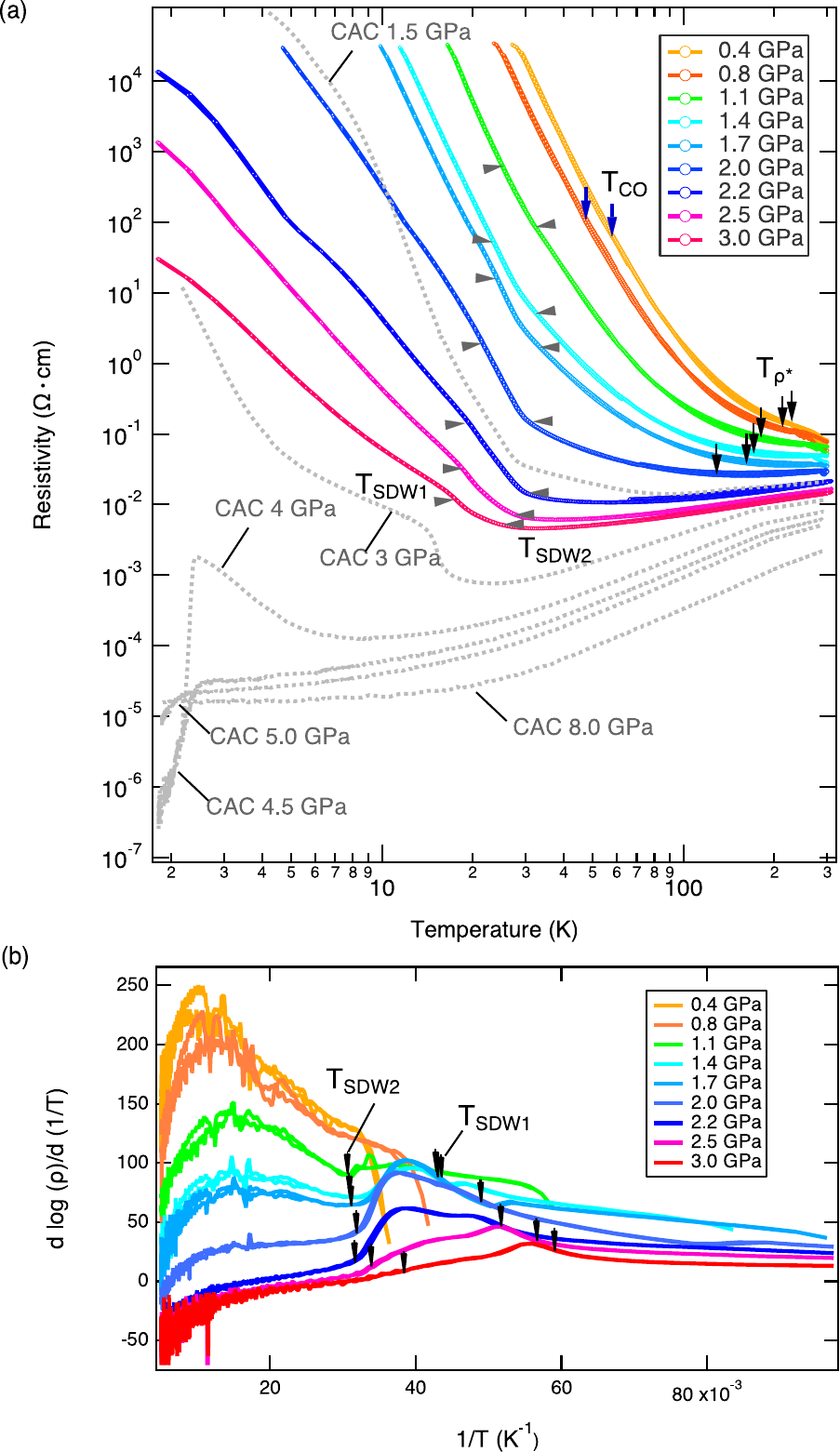}
\caption{
The temperature dependence of resistivity $\rho(T)$ of (TMTTF)$_2$PF$_6$, 
under pressures from 0.3 to 3.0~GPa (a) in the log-log plot and (b) in the form of $d(\log \rho)/d(1/T)$ vs $1/T$. 
In (a), the results for higher pressure obtained by a cubic anvil cell (CAC)~\cite{araki2007electrical, kano2012anisotropy} are also shown by the gray dotted lines, showing superconducting behavior at around 4~GPa. 
The characteristic temperatures found in the plot in (a): $T_{\rho}^*$ and $T_{\rm{CO}}$, and in (b): $T_{\rm{SDW1}}$ and $T_{\rm{SDW2}}$ as defined in the text are shown by arrows.}

\label{fig-resistivity}
\end{center}
\end{figure}

Fig.~\ref{fig-resistivity}(a) shows the temperature dependence of the resistivity of (TMTTF)$_2$PF$_6$ at different pressures, displayed in the log-log plot; data for $0.4 < P < 3.0$~GPa are shown together with the previously obtained data using a cubic anvil cell (CAC) up to 8~GPa~\cite{araki2007electrical, kano2012anisotropy}. 
Although there are minor differences due to the different pressure equipment setups, the behavior is consistent with previous reports~\cite{Moser, Adachi_doi:10.1021/ja0001254, DJaccard_2001}.
Let us analyze them in the following, by comparing to previous reports and other measurements such as NMR which has been used to track the phase evolution under pressure microscopically.

At ambient pressure, from room temperature the resistivity increases upon cooling, which is attributed to the DM insulator, 
 and the CO phase transition takes place at $T_{\rm{CO}} \sim$ 
 70~K~\cite{Chow_2000,CO-PhysRevLett.86.4080}. 
The phase transition was known as the ``structureless anomaly",  
 where the emergence of charge disproportionation was detected in NMR~\cite{Chow_2000} and dielectric~\cite{CO-PhysRevLett.86.4080} measurements. 
 While neutron scattering suggested a slight structural displacement at the CO transition~\cite{Newtron_FOURYLEYLEKIAN2010S95}, 
 a recent synchrotron x-ray diffraction study revealed the difference 
 of the two TMTTF molecules in the dimer owing to CO~\cite{Kitou_PhysRevLett.119.065701}.
By applying pressure the resistivity monotonically decreases and at low pressures of about $P < 1$~GPa similar behavior is observed. 
Eventually, a resistivity minimum $T_{\rho}$ appears and shifts to lower temperature as pressure is increased; 
 This behavior is interpreted as a metal-insulator crossover owing to the DM gap~\cite{Emery_PRL1982}.
On the other hand, the CO transition in (TMTTF)$_2$PF$_6$ does not clearly appear in the resistivity~\cite{Kohler_PhysRevB.84.035124} 
at ambient pressure, possibly related to the freezing temperature of rotating PF$_6$ anions~\cite{Yu_PRB2004}.

Here, we identify the characteristic temperatures as follows:
First, at low pressure below 2~GPa, the slope changes 
 at around 100 K -- 200 K [see Fig.~\ref{fig-resistivity}(a)]. 
In the $a$-axis resistivity at ambient pressure, a minimum ($T_{\rho}$) from a metallic state on the high temperature side appears~\cite{Kohler_PhysRevB.84.035124}.
However, in our result here, the resistivity increases with decreasing temperature from room temperature, and its slope changes at a certain temperature, which we denote as $T_{\rho}^*$;
 we consider this to be the same origin as $T_{\rho}$, i.e., the development of the Mott gap, 
 judged from their similar pressure dependencies ~\cite{Moser}. 

Next, we attribute the CO transition temperature 
 by referring to the ambient pressure value $T_{\rm CO} =$~70~K. 
The $T_{\rm CO}$ in (TMTTF)$_2$PF$_6$ appears slight change in $\rho$ vs $T$ as shown in Fig.~\ref{fig-resistivity}(a), 
 which is less clear than $X$ = AsF$_6$ and SbF$_6$ compounds~\cite{Kohler_PhysRevB.84.035124}.
$T_{\rm CO}$ shifts to lower temperature with increasing the pressure, and disappears at around 1~GPa. 
 
Finally, we ascribe  
the development of the SDW state 
 in the $d(\log \rho)/d(1/T)$ vs $1/T$ plot, displayed in Fig.~\ref{fig-resistivity}(b). 
Above 1~GPa, a peak appears at around 30~K, which becomes pronounced as the pressure is increased. 
We defined the bases of the peak as $T_{\rm SDW1}$ (lower temperature) and $T_{\rm SDW2}$ (higher temperature).
$T_{\rm SDW2}$ are higher than the transition temperatures in previous reports of the resistivity~\cite{Moser} and NMR measurements \cite{Chow_PhysRevLett.81.3984}, 
which might be owing to the quasi-one-dimensional nature of the phase transition; $T_{\rm SDW2}$ is where the gap starts to open along the TMTTF chains.

\begin{figure}
\begin{center} 
\includegraphics[width=80mm]{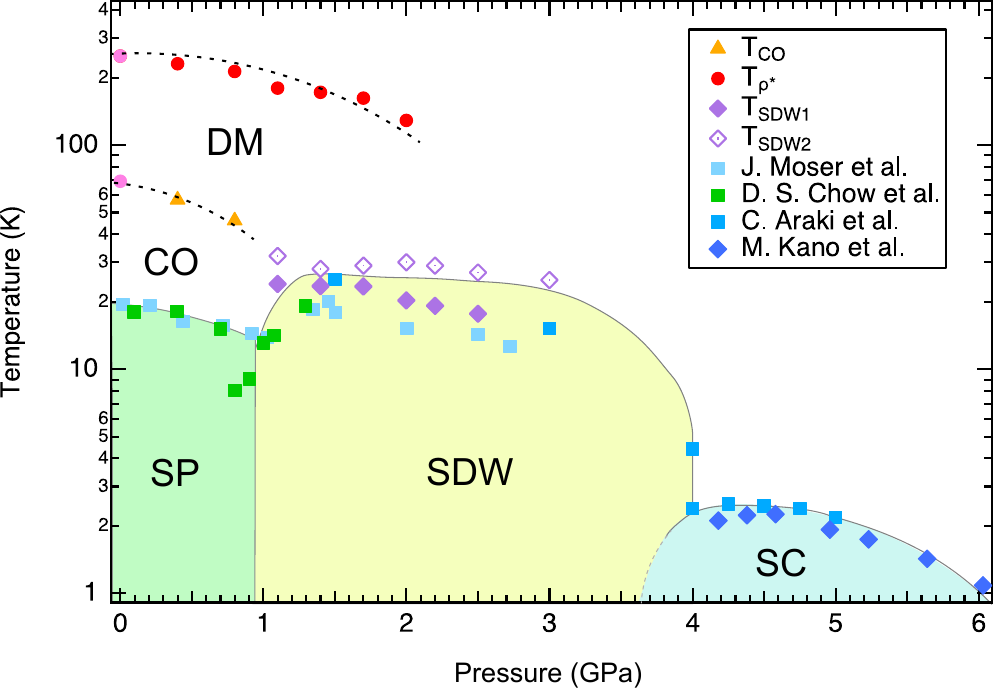}
\caption{Experimental pressure-temperature phase diagram 
of (TMTTF)$_2$PF$_6$. DM, CO, SP, SDW, and SC refers to the dimer-Mott insulating state, charge order, spin-Peierls, spin-density-wave, and superconducting phases, respectively. 
Crossover temperature to the DM state, $T_{\rho}^*$, and transition temperatures to CO and SDW, $T_{\rm{CO}}$ and $T_{\rm{SDW}}$, respectively,
obtained by our resistivity measurement up to 3 GPa, are plotted together with the data points from previous works~\cite{Chow_PhysRevLett.81.3984,Chow_2000,Moser,araki2007electrical,kano2012anisotropy}. The transition temperature of DM and CO at ambient pressure (dark pink circles) are referred to \cite{Kohler_PhysRevB.84.035124}}. 
\label{P-T}
\end{center}
\end{figure}

The phase transition from the SDW state to a superconducting state proceeds by an external pressure above 4~GPa, which is not produced by a usual clamp-type pressure cell. 
By the result of resistivity measurement using CAC above 1.8~K, the superconducting phase exists in the pressure between 4~GPa and 5~GPa, and the maximum $T_{\rm{C}}$ is 2.5~K at 4.3~GPa \cite{araki2007electrical}. 
In the case of the result using DAC, the observed superconducting phase is in 4.18 GPa $\leq P \leq$ 6.03 GPa and the maximum $T_{\rm{C}}$ is 2.25~K at 4.58 GPa \cite{kano2012anisotropy}.
Figure~\ref{P-T} summarizes the pressure-temperature diagram of (TMTTF)$_2$PF$_6$, in which the crossover and transition temperatures obtained by our resistivity measurement are added to the phase boundaries in previous reports~\cite{Chow_2000, Moser,kano2012anisotropy,araki2007electrical, Chow_PhysRevLett.81.3984}.

\subsection{Crystal structure}\label{subsec:strucure}

\begin{figure}
\begin{center} 
\includegraphics[width=80mm]{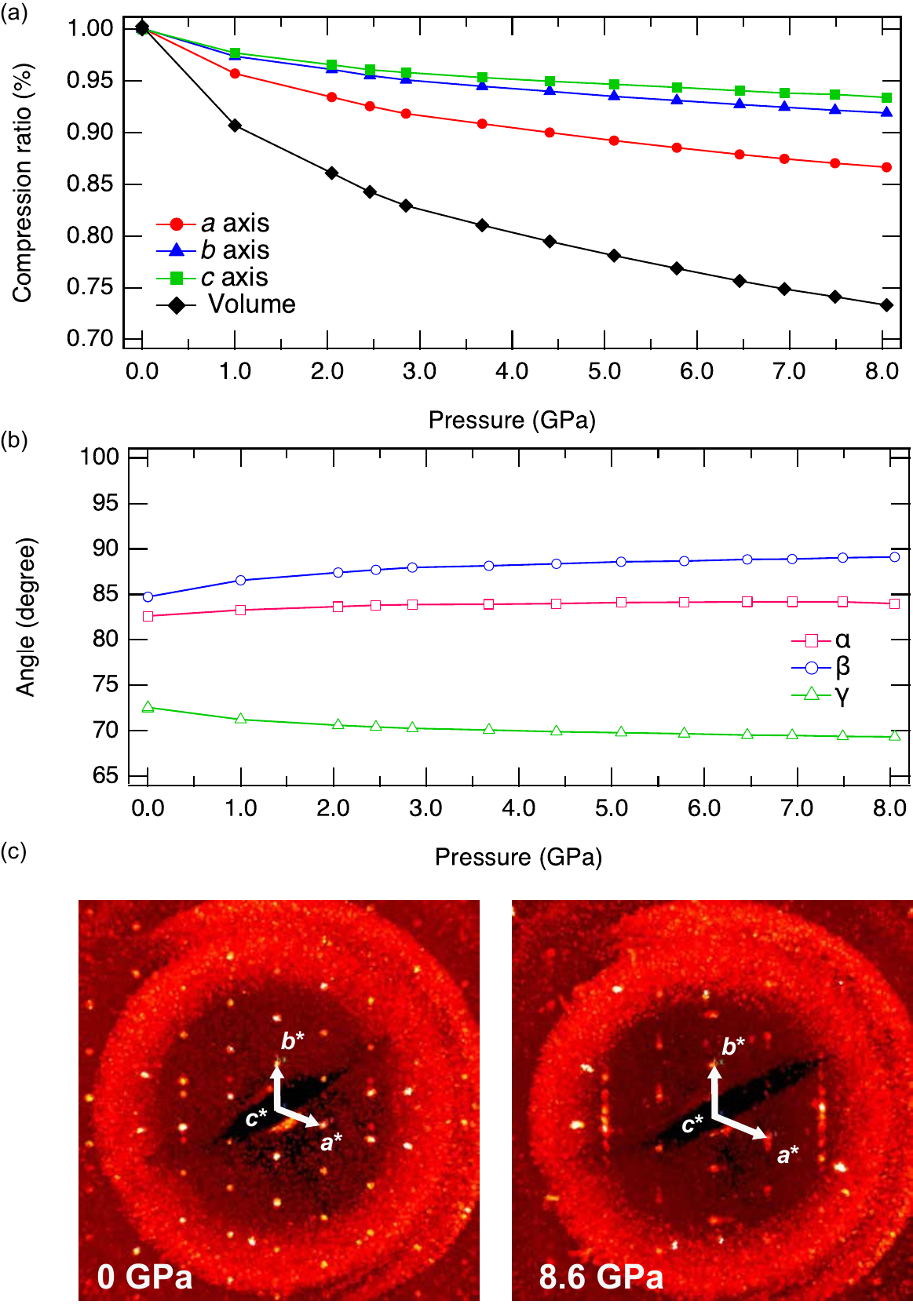}
\caption{Pressure dependence of (a) compression ratio of the lattice parameters and volume with respect to their ambient pressure values, and (b) unit cell angles, up to 8~GPa in which the space group 
remains $P\bar{1}$.  
(c) Diffraction spots observed from $a$* $b$* plane under 0~GPa and 8.6~GPa. The white ring outside the clear diffraction spot of (TMTTF)$_2$PF$_6$ is diffraction from the Re gasket.}
\label{fig-lattice}
\end{center}
\end{figure}

The lattice parameters of (TMTTF)$_2$PF$_6$ obtained by the single crystal
XRD measurements up to 8.1~GPa are listed in Table~\ref{tab:Lattice parameters} (Appendix A), whose data are depicted in Fig. \ref{fig-lattice}(a) and \ref{fig-lattice}(b), as the lattice compression ratios and angles, respectively. 
The crystal symmetry remains $P\bar{1}$ up to 8~GPa, whereas a structural transition occurs at 8.5~GPa, above which the diffraction pattern clearly changes as shown in Fig.~\ref{fig-lattice}(c), which is reported as a triclinic to monoclinic transition~\cite{pashkin2009pressure}.
Once the crystal was compressed above 8.5~GPa, the crystal packing never returned to its original state after the pressure was released.
Up to about 3~GPa, the unit cell volume ($V$) is rapidly compressed down to 80 $\%$ of that at ambient pressure (see Fig.~\ref{fig-lattice}(a). Above 3~GPa, the $V$ gradually shrinks and becomes 70 $\%$ at 8~GPa.

Bulk modulus $B_0$ and its derivative $B'_0$ are evaluated by fitting of $V/V_0$ vs $P$ curve with the 
Birch-Murnaghan equation of state \cite{Birch_PhysRev.71.809}. 
The obtained $B_0$ and $B'_0$ are 6.5(2)~GPa and 10.6(5), which are comparable to those reported in reference \cite{pashkin2009pressure}. 
The value of the bulk modulus is almost half value of (TMTSF)$_2$PF$_6$ ($B_0$ = 12.7(9)~GPa \cite{pashkin2009pressure}), whose the superconducting phase appears at 0.9 K and 1.2 GPa \cite{jerome1980superconductivity}. 
It should be noted that the (TMTTF)$_2$PF$_6$ is quite soft compared to typical coordination polymers such as Prussian blue analogues RbCo[Fe(CN)$_6$] and RbNi[Cr(CN)$_6$] ($B_0$ = 25(1) and 19(2)~GPa, $B'_0$ = 29.0(5) and 20(1), respectively)~\cite{PBA_10.1063/5.0049223}. 

As seen in Fig.~\ref{fig-lattice}(a), for the whole pressure region, the $a$ axis is more compressed by external pressure ($\sim$ 10 $\%$) than the $b$- and $c$-axes: Their compression ratios are almost the half of that of the $a$ axis ($\sim$ 5 $\%$ for $b$ and $c$ axes). 
This is because the PF$_6$ anions prevent the compression between the TMTTF chains along the $a$ axis. 
In the process of compression, while the angle $\alpha$ is almost unchanged,  $\beta$ gradually increases toward 90$^{\circ}$, whereas $\gamma$ gradually decreases, whose variations are limited to low pressures up to about 3--4~GPa. 
These features are consistent with the previous reports~\cite{pashkin2009pressure, Rose_2013}.

\begin{figure}
\begin{center} 
\includegraphics[width=85mm]{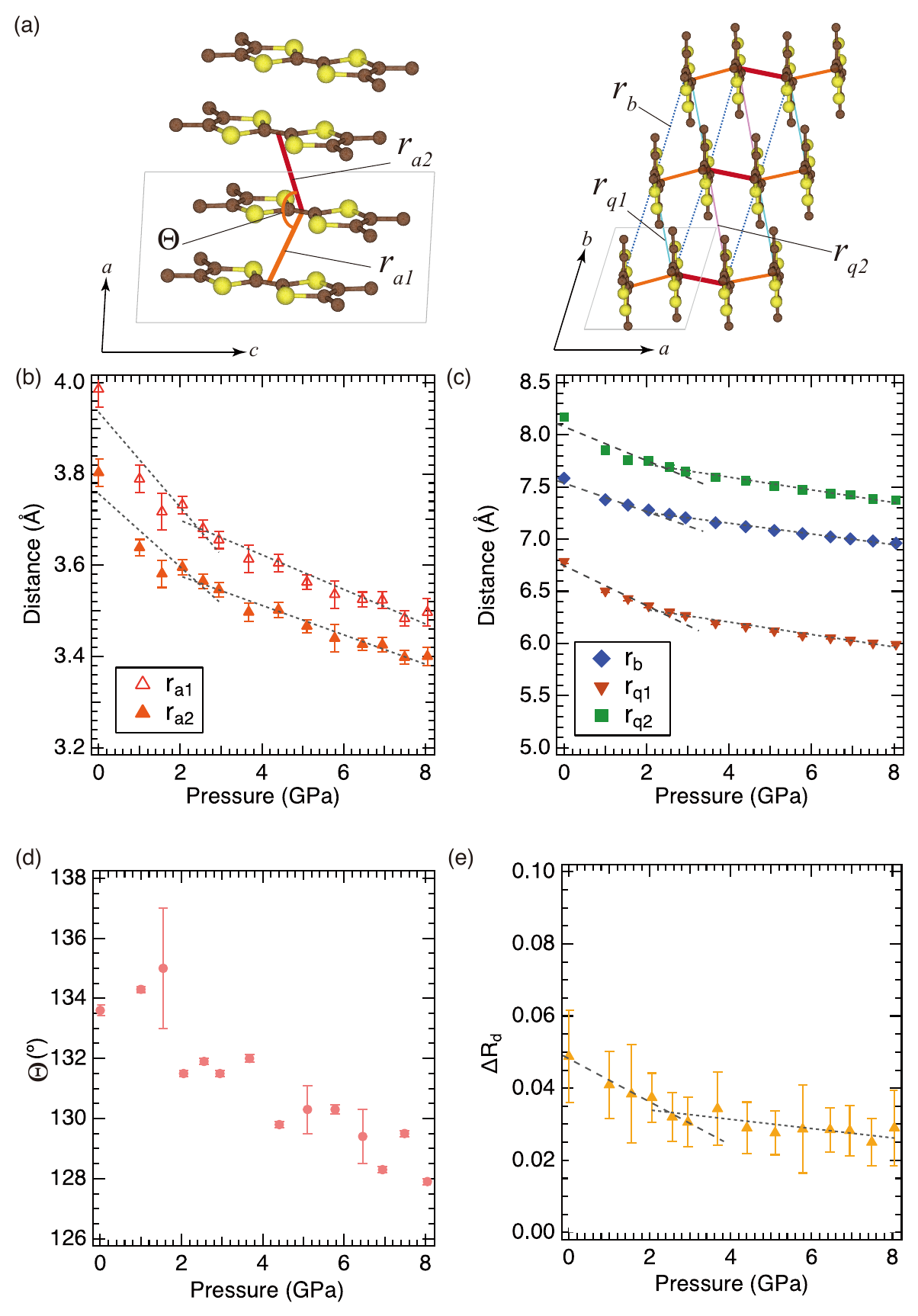}
\vspace{-0.5cm} 
\caption{(a) Definition of the intermolecular distances between TMTTFs and the shift angle along the chain direction.
Pressure dependencies of the distance between TMTTF molecules for the (b) intrachain and (c) interchain directions, (d) the shift angle $\Theta$, and (e) the relative dimerization $\Delta R_d$. For their definitions, see the text. Dash lines and dot lines are obtained by liner regression and show that the inclinations of intermolecular and intrachain distances changes between 2 and 3 GPa.}  
\label{fig-ta}
\end{center}
\end{figure}

From the full structural parameters, we estimated the distances between the adjacent TMTTF molecules depicted in Fig.~\ref{fig-ta}(a), by using the middle coordinates of the central C=C bonds. 
Figures~\ref{fig-ta}(b)~and~\ref{fig-ta}(c) show the 
pressure dependencies of the intrachain ($r_{a1}$ and $r_{a2}$) and interchain ($r_b$, $r_{q1}$, and $r_{q2}$) distances, all showing gradual decrease with increasing pressure.
It should be noted that $r_{a1}$ (12 \%) and  $r_{a2}$ (11 \%) show a higher compression ratio than other distances (less than 10 \%); 
these are related to the pressure dependencies of the model parameters discussed in Sec. III. 

We also plot the shift angle ($\Theta$) between TMTTF molecules along $a$ axis, defined in Fig.~\ref{fig-ta}(a), in Fig.~\ref{fig-ta}(d) which decreases monotonically by applying pressure. 
This decrease indicates that the TMTTF molecules move in the interchain and interlayer directions without an obvious development of the zigzag arrangement, since the bulky anions are situated in between the molecules. 
Furthermore, we estimate the degree of structural dimerization as $\Delta R_d = 2(r_{a2}-r_{a1})/(r_{a1}+r_{a2})$, plotted in Fig.~\ref{fig-ta}~(e).
It decreases up to 3 GPa, and above that pressure, $\Delta R_d$ remains almost constant. 

Using first-principles methods, we have also obtained the parameters by structural optimization for the internal atomic coordinates, which show similar tendencies but small differences are seen; the structures are compared in Appendix B.

\section{Theoretical analysis}\label{sec:theory}

\subsection{Methods}\label{subsec:method}

First, using the experimental room temperature crystal structures under pressure, we perform density functional theory (DFT) calculations by \texttt{Quantum Espresso (version 6.6)}~\cite{QE}. 
The positions of the F and H atoms forming PF$_6$ and CH$_3$ in TMTTF molecules were relaxed because the forces acting on them were excessive within the experimentally determined structures (see Appendix B). 
We choose methods that are commonly used for describing electronic states of molecular compounds~\cite{Ishibashi-2009}: Norm-conserving pseudopotentials based on the Vanderbilt formalism with plane-wave basis sets~\cite{Hamann_ONCV2013, Schlipf_CPC2015}, and the generalized gradient approximation by Perdew, Burke, and Ernzerhof~\cite{GGA_PBE} as the exchange-correlation functional.
We derive maximally localized Wannier functions (MLWFs) from the Kohn-Sham orbitals. 
The cutoff energies for plane waves and charge densities are set to $70$ and $280$ Ry, respectively. 
A Gaussian smearing method was used with a $7\times 7\times 3$ uniform $\bm{k}$-point mesh during the self-consistent loops.
From the obtained transfer integrals between the MLWFs, the Fermi surfaces and the bare irreducible susceptibilities are obtained using \texttt{H-wave}~\cite{AOYAMA2024109087}. 
For the latter, we set the lattice size to $N_x = N_y = 64$ and $N_z = 8$, temperature to $T = 0.01$ (eV), and the range of fermionic Matsubara frequencies as $-\pi T(N_c + 1), \cdots, \pi T (N_c - 1)$ with $N_c = 1024$.

Following the DFT calculations, we evaluate the interaction parameters within the constrained random phase approximation (cRPA)~\cite{PhysRevB.70.195104, Imada_JPSJ2010}, using \texttt{RESPACK}~\cite{RESPACK}. 
We note that previous studies applying the cRPA method to molecular solids show good agreement between the results for the derived effective Hamiltonians and the experiments ~\cite{Shinaoka2012, PhysRevResearch.2.032072, PhysRevResearch.3.043224, Ido2022, Ohki2023}, which has recently adopted to the TM salts~\cite{yoshimi2022}. 
Then the following {\it ab initio} extended Hubbard Hamiltonian in two dimensions is obtained:
\begin{align}
&H=
\sum_{ij,\sigma}t_{ij}(c_{i\sigma}^{\dagger}c_{j\sigma}+{\rm h.c.})
+U\sum_{i}n_{i\uparrow}n_{i\downarrow}
+\sum_{ij}V_{ij}N_{i}N_{j},
\label{eq:Ham}
\end{align}
where $c^{\dagger}_{i\sigma}$ and $c_{i\sigma}$ are the creation and annihilation operators of an electron with spin $\sigma$ at the $i$-th site, respectively.
The number operators are defined as $n_{i\sigma}=c_{i\sigma}^{\dagger}c_{i\sigma}$ and
$N_{i}=n_{i\uparrow}+n_{i\downarrow}$.

We solve the $ab$ $initio$ model in Eq.~(\ref{eq:Ham}) using the many-variable variational Monte Carlo (mVMC) method using \texttt{mVMC}\cite{misawa_CPC2019}. 
In the mVMC analysis, to take into account 
inter-layer screening effects, we have subtracted a constant value $\Delta_{\rm DDF}=0.2$ eV from the effective Coulomb interactions,
following previous studies~\cite{Nakamura_JPSJ2010, PhysRevB.86.205117}.
We employ the following trial wave functions:
\begin{align}
\ket{\psi}&=\mathcal{L}_{S}\mathcal{P}_{J}\mathcal{P}_{G}\ket{\phi_{\rm pair}},\\
\ket{\phi_{\rm pair}}&=\Big(\sum_{i,j}f_{ij}c_{i\uparrow}^{\dagger}c_{j\downarrow}^{\dagger}\Big)^{\frac{N_{\rm e}}{2}}\ket{0},
\end{align}
where $\mathcal{L}_{S}$, $P_{G}$, and $P_{J}$ represent
the total spin projection, the Gutzwiller factor, and the Jastrow factor, respectively. 
We note that the pair-product part $\ket{\phi_{\rm pair}}$ 
can describe a wide range of quantum phases, such as spin/charge ordered phases, 
superconducting phase, and quantum spin liquids.
In accordance with the previous study~\cite{yoshimi2022}, we use the spin singlet projection and impose a $4\times2$ sublattice structure to express the spin/charge ordered phases in the TM salts. 
We optimize all the variational parameters simultaneously using the stochastic reconfiguration method~\cite{Sorella_PRB2001} 
to obtain the ground state.

\begin{figure}[b] 
\begin{center} 
\includegraphics[width=0.95\columnwidth]{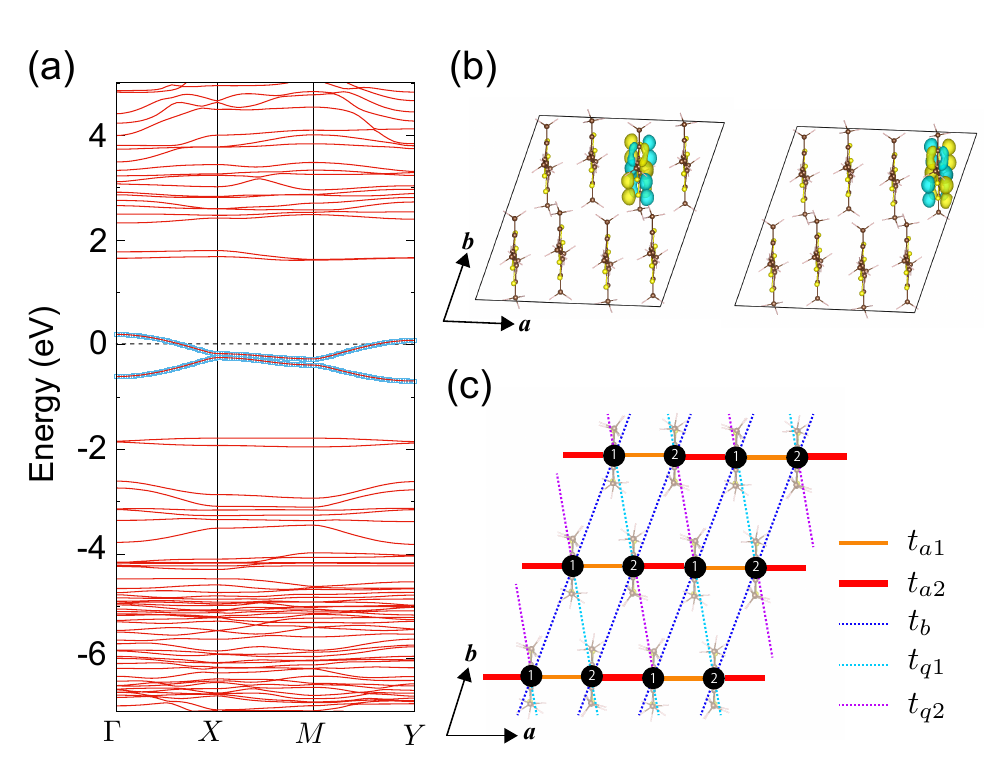}
\caption{
(a) Band structure of (TMTTF)$_2$PF$_6$ at ambient pressure. The thin solid lines are the DFT results, while the bold lines are the tight-binding bands obtained using the MLWFs, overlapping the valence bands. 
We set the Fermi energy to zero (the dotted line). 
Here, $\Gamma = (0, 0, 0), X = (\pi, 0, 0), M = (\pi, \pi, 0),$ and $Y=(0, \pi, 0)$.
(b) Drawings of the two MLWFs for (TMTTF)$_2$PF$_6$ . 
(c) Definition of the transfer integrals. 
The crystal structure and the MLWFs are drawn using \texttt{VESTA}~\cite{VESTA}.} 
\label{fig-bands} 
\end{center}
\end{figure}

\subsection{Electronic structure}\label{subsec:electronic_strucuture}

\begin{figure}
    \centering
    \includegraphics[width=60mm]{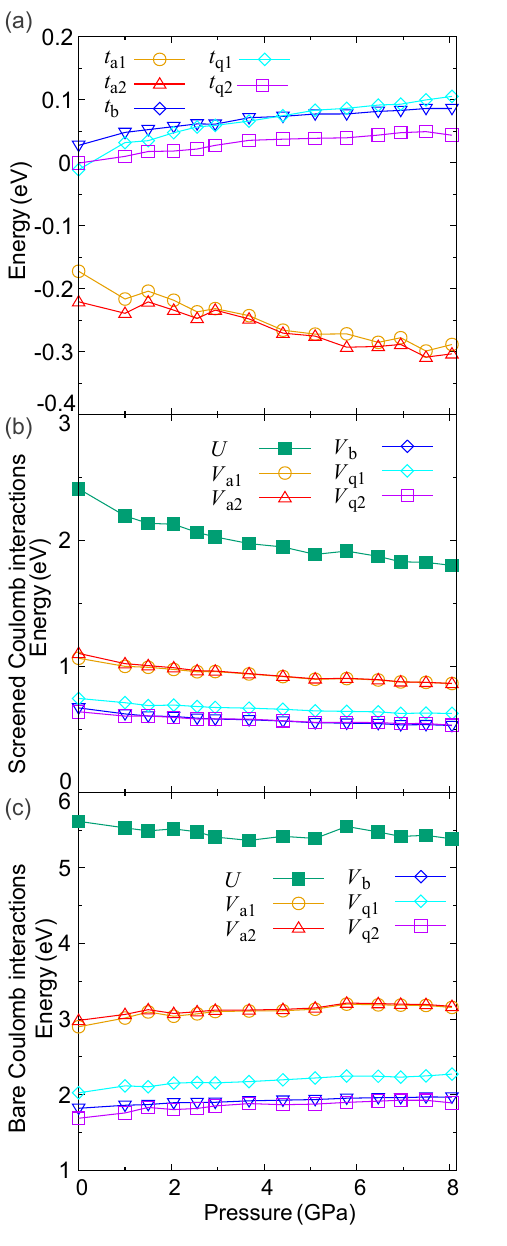}
    \caption{Pressure dependence of (a) the transfer integrals, (b) the screened and (c) the bare Coulomb interactions, obtained by using the MLWFs obtained by the DFT calculations.}
    \label{fig:transfer_and_V}
\end{figure}
Figure
~\ref{fig-bands}(a) shows the band structure at ambient pressure.
The isolated two bands near the Fermi energy are constituted from the bonding and anti-bonding combination of the MLWFs located on the two molecules in the unit cell shown in Fig.~\ref{fig-bands}(b). 
As mentioned above, using these MLWFs, we evaluate the transfer integrals and the density-density interactions, i.e. the on-site and off-site Coulomb interactions (the exchange interactions are negligibly small, so we omit them here). 

The obtained microscopic parameters are summarized in Fig.~\ref{fig:transfer_and_V}.
Following previous studies~\cite{Yoshimi_2012,Yoshimi2012_PhysicaB, yoshimi2022,note_bondnotation}, we adopt the bond indices as shown in Fig. \ref{fig-bands}(c). 
A general trend is that, first, the absolute values of the transfer integrals monotonically increase by pressure [Fig.~\ref{fig:transfer_and_V}(a)], as expected 
 from the compression of the crystal leading to the increase of inter-molecular distances [see Figs.~\ref{fig-ta}(b) and \ref{fig-ta}(c)]. 
 On the other hand, the interaction parameters show the opposite trend, as shown in Fig.~\ref{fig:transfer_and_V}(b); 
  this is owing to the screening effect considered in the cRPA method, by the increase of bandwidths in the bands other than the target bands. 
  In fact, the bare value of $U$ shows only a small change by the pressure, and those of the inter-site terms indeed increase as shown in Fig.~\ref{fig:transfer_and_V}(c). 
We emphasize that such a trend under pressure has been directly demonstrated for the first time, to the best of the authors' knowledge.

\begin{figure}
\begin{center} 
\includegraphics[width=75mm]{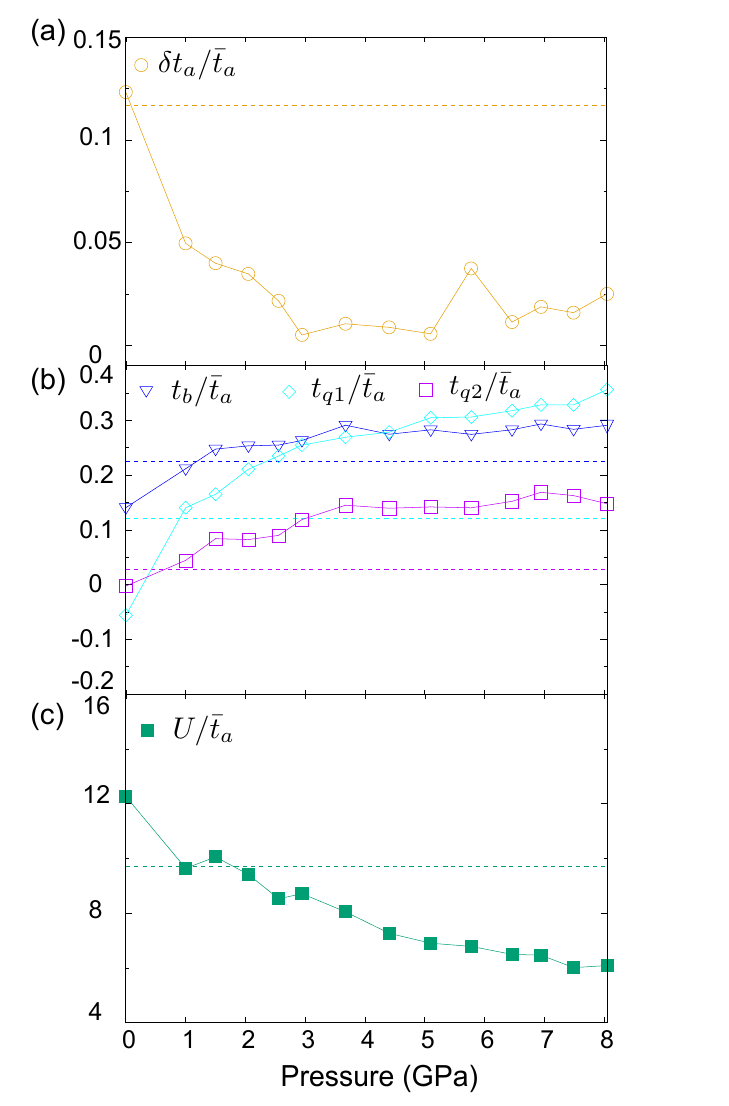}
\vspace{-0.5cm} 
\caption{
(a) Difference in the alternating transfer integrals along $a$ axis ($\delta t_a \equiv |t_{a2}-t_{a1}|$), (b) $t_b$, $t_{q1}$, $t_{q2}$, and (c) the on-site Coulomb interaction ($U$) for (TMTTF)$_2$PF$_6$ (square symbols) and (TMTSF)$_2$PF$_6$ (dotted lines)\cite{yoshimi2022}. Each value is normalized by the mean $a$ axis transfer integral, $\bar{t}_a\equiv |(t_{a1}+t_{a2})/2|$.} 
\label{fig-params}
\end{center}
\end{figure}

To discuss the effect of pressure on the electronic structure, we extract several parameters.  
Figure \ref{fig-params}(a) shows the degree of dimerization in the chain direction, $\delta t_a/\bar{t}_a$, where the difference $\delta t_a \equiv |t_{a2}-t_{a1}|$ is normalized by the mean value $\bar{t}_a = |(t_{a1} + t_{a2})/2|$. 
As the pressure increases, surprisingly, $\delta t_a/\bar{t}_a$ rapidly becomes small and reaches almost $0$ above $P = 3$ GPa. 
In contrast, the interchain transfer integrals (normalized by $\bar{t}_a$), plotted in Fig. \ref{fig-params}(b), i.e.,  
$t_b/\bar{t}_a$, $t_{q1}/\bar{t}_a$, and $t_{q2}/\bar{t}_a$, monotonically increase. 
A notable point is the large increase in $t_{q1}/\bar{t}_a$ compared to the others.
Figure \ref{fig-params} (c) shows the normalized on-site Coulomb interaction $U/\bar{t}_a$. 
The pressure severely affects the value of $U/\bar{t}_a$, which is at $8$ GPa almost half as small as that at ambient pressure.

For comparison, the values for (TMTSF)$_2$PF$_6$ at 0~GPa~\cite{yoshimi2022}, that situates near the right end of the unified phase diagram, are shown as dotted lines. 
If we focus on Figs. \ref{fig-params}(b) and \ref{fig-params}(c), the chemical pressure from (TMTTF)$_2$PF$_6$ to 
(TMTSF)$_2$PF$_6$ corresponds to an applied pressure of around 1.5~GPa. 
However, in Fig. \ref{fig-params}(a), 
the degree of dimerization between the two is similar at ambient pressure, and by applying pressure to (TMTTF)$_2$PF$_6$ it rapidly decreases as mentioned above. 
This indicates a clear difference between the physical and chemical pressure effects.

\begin{figure*}
\begin{center} 
\includegraphics[width=2.2\columnwidth]
{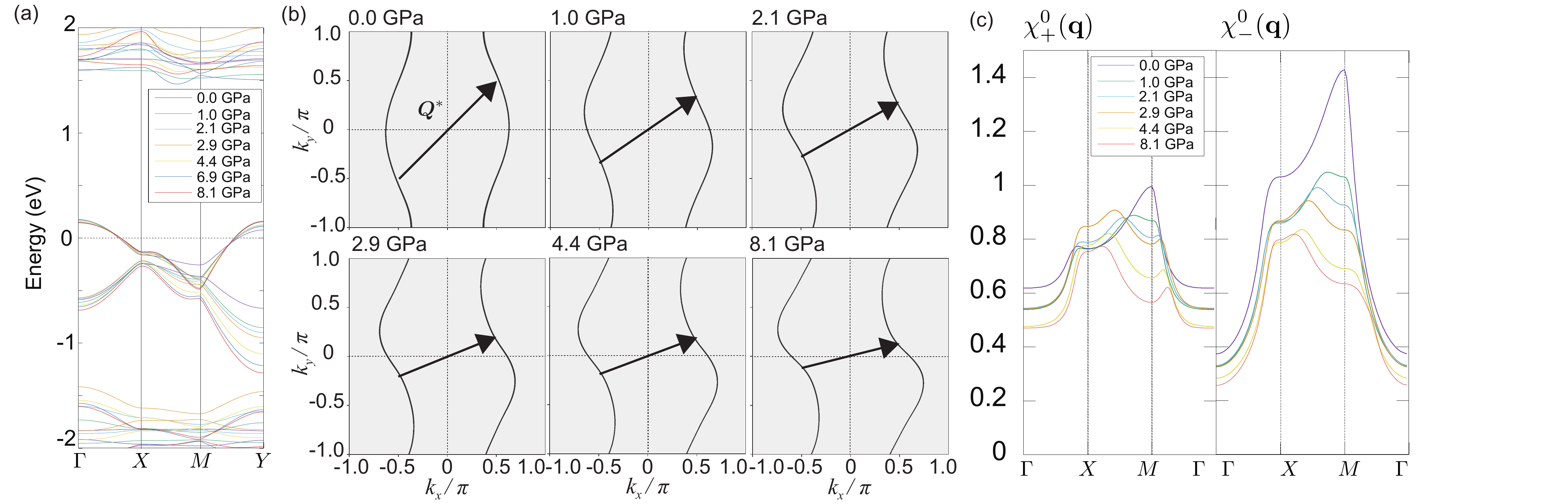}
\caption{(a) Band structures, (b) Fermi surfaces, obtained from first-principles DFT calculations, and (c) irreducible susceptibilities $\chi_{\pm}^{0}({\bm q})$, defined in the text, of the tight-binding model for (TMTTF)$_2$PF$_6$ at different pressures.
Here, $\Gamma = (0, 0, 0), X = (\pi, 0, 0), M = (\pi, \pi, 0),$ and $Y = (0, \pi, 0)$.
The nesting vectors $\bm{Q}^* =(\pi, q_y^{\rm max}, *)$ are about $q_y^{\rm max} = \pi, 9/16\pi, 13/32\pi, 3/8\pi, \pi/4$ at $P=0.0$, $1.0$, $2.1$, $3.0$, $4.4$, and $8.1$~GPa, respectively.
} 
\label{fig-FS-p}
\end{center}
\end{figure*}
Figure ~\ref{fig-FS-p}(a) shows the change in the band structure by applying pressure.
As $P$ increases, the bandwidth increases whereas the dimerization gap, i.e., the gap between the two bands at the $X$ point, is almost unchanged. 
Although the in-chain dimerization $\delta t_a/\bar{t}_a$ decreases down to almost 0 as we mentioned above,  
 at high pressure the gap reflects the difference between $t_{q1}$ and $t_{q2}$. 
The increase of the bandwidth in the interchain direction is pronounced, 
which is also consistent with the trend in the interchain parameters.
In Fig.~\ref{fig-FS-p}(b), the Fermi surfaces at different pressures are shown. 
One can see that the shape of the Fermi surfaces is open and warped throughout the whole pressure range,
and as pressure is applied, it becomes more asymmetric with respect to the $k_y$ axis up to about 2 GPa, and then more or less unchanged above that pressure. This asymmetry is owing to the increase in $t_{q1}$ and $t_{q2}$. 

The nesting vector is an important quantity in the instability towards density-wave states, and associated with the periodicity of the SDW state in this system.
We can detect it by the peak(s) in the bare irreducible susceptibility. 
Figure~\ref{fig-FS-p}(c) shows the momentum (${\bm q}$) dependent susceptibilities $\chi_{\pm}^{0}({\bm q})$, defined as the combinations $\chi_{\pm}^0({\bm q}) \equiv \chi_{11}^0({\bm q})\pm \chi_{12}^0({\bm q})$ where $\chi_{\alpha \beta}^0({\bm q})$ are the matrix elements in the site representation, for different pressures. Here, $\chi_{+(-)}^0({\bm q})$ corresponds to the inter(intra)-dimer fluctuation in the non-interacting system.
It can be seen that there is a peak near ${\bm Q}^* = (\pi, \pi, \ast)$ (the peak does not depend on $q_z$) at ambient pressure, and changes to an incommensurate value in the $y$-direction as ${\bm Q}^* = (\pi, q^{\rm max}_y, \ast)$. 
We show the nesting vector ${\bm Q}^*$ where $\chi_-^0({\bm q})$ has the maximum value in Fig. \ref{fig-FS-p} (b). 
Considering previous studies for (TMTSF)$_2$PF$_6$ indicating an incommensurate nesting vector to be close to $(\pi, \pi/2, *)$~\cite{Jerome_ChemRev}, the periodicity of the SDW state under pressure in (TMTTF)$_2$PF$_6$ discussed in Sec.~\ref{experiments} 
is suggested to be close to that in (TMTSF)$_2$PF$_6$.

\subsection{mVMC analysis}~\label{subsec:mVMC}

\begin{figure*}
\begin{center} 
\includegraphics[width=1.8\columnwidth]{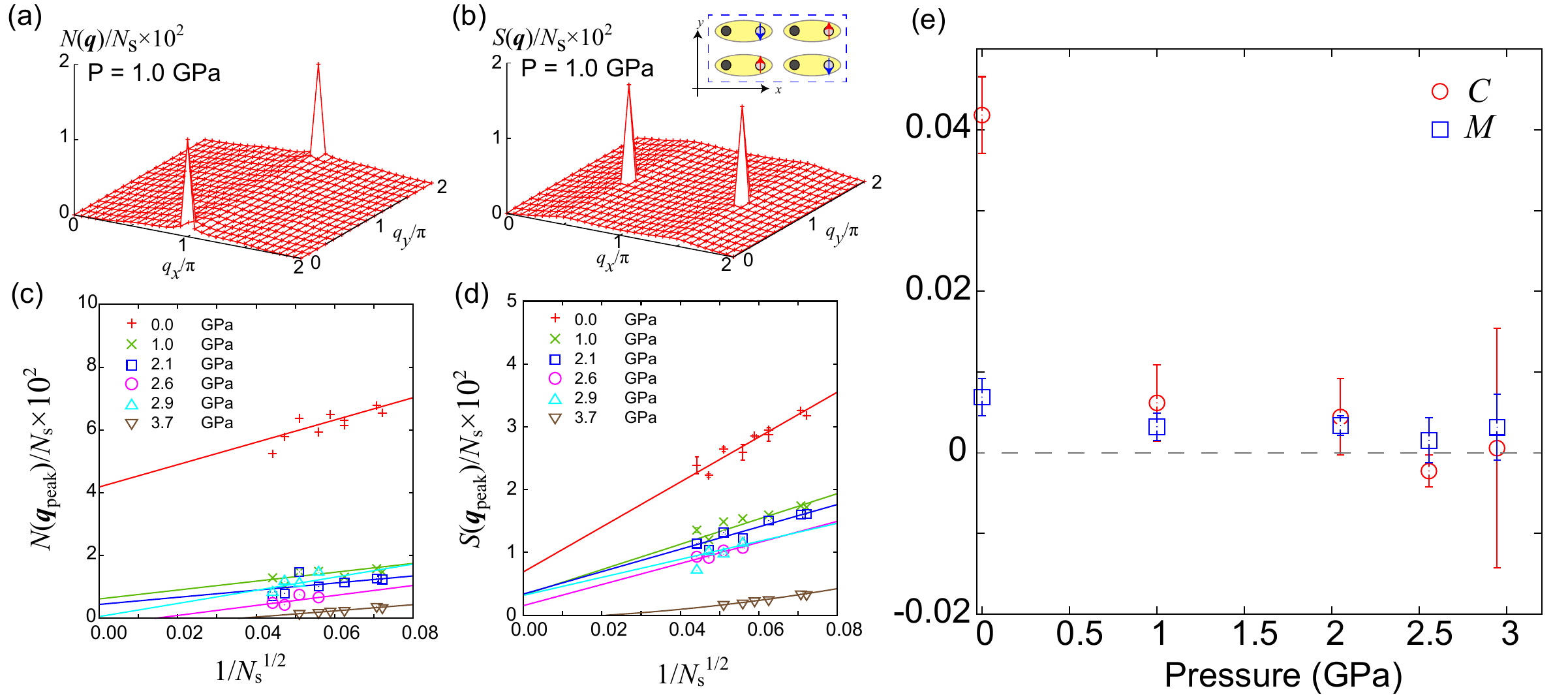}
\caption{(a) Charge [$N(\vec{q})$] and (b) spin [$S(\vec{q})$] structure factors 
at 1~GPa for $L_x=32, L_y=16$. 
Size dependence of the peak value of (c) charge and (d) spin structure factors for several pressures.
(e) Pressure dependence of the order parameters of the charge and spin orderings.} 
\label{fig:NqSq}
\end{center}
\end{figure*}

Finally, we show the results of the mVMC calculations of the two-dimensional {\it ab initio} model in Eq.~(\ref{eq:Ham}), for the parameter sets at 6 different pressures (including the ambient pressure) up to those at 3.7~GPa.
In Figs.~\ref{fig:NqSq}(a) and \ref{fig:NqSq}(b),  charge and spin structure factors which are defined as
\begin{align}
N(\vec{q})&=\frac{1}{N_{\rm s}}\sum_{i,j}\ev*{\bar{N}_{i}\cdot \bar{N}_{j}}e^{i\vec{q}\cdot(\vec{r}_{i}-\vec{r}_{j})}, \label{eq:Nq}\\
S(\vec{q})&=\frac{1}{N_{\rm s}}\sum_{i,j}\ev*{\vec{S}_{i}\cdot\vec{S}_{j}}e^{i\vec{q}\cdot(\vec{r}_{i}-\vec{r}_{j})}, 
\end{align}
are shown, for the parameters at 1~GPa. Here, $\bar{N}_{i}=N_{i}-\ev*{N}$ and $\ev*{N}=(1/N_{\rm s})\sum_{i}N_{i}$. 
Note that we map the original lattice with a dimer structure onto the $N_{\rm s}=L_{x}\times L_{y}$ lattice. 
We find that the charge (spin) structure factors have a peak at $\vec{q}_{\rm peak}=(\pi,0)$ [$\vec{q}_{\rm peak}=(\pi/2,\pi)$], whose schematic picture of the corresponding ordered phase is shown in the inset of Fig.~\ref{fig:NqSq}(b). 
These features are similar to our previous results at ambient pressure~\cite{yoshimi2022}, which we have quantitatively confirmed as the same. This confirmation is achieved by performing the mVMC calculations for the effective Hamiltonians derived from the structural data obtained in this study.
However, the peak values are reduced in the 1~GPa case. The peak in the charge (spin) structure factor is about 20\% (50\%) of its ambient pressure value [see Figs.~\ref{fig:NqSq}(c) and \ref{fig:NqSq}(d)]; the reduction is more pronounced in the charge sector.

We show size dependence of $N(\vec{q}_{\rm peak})$ and $S(\vec{q}_{\rm peak})$
for different pressures in Figs.~\ref{fig:NqSq}(c) and (d).
By increasing pressure, the peak value of the charge structure factors suddenly decreases while
that of the spin structure factors gradually decreases. 
Although the size dependence of $N(\vec{q}_{\rm peak})$ is not smooth, 
it is probable that the CO almost vanishes in the thermodynamic limit for $P\geq 1$~GPa.
On the other hand, the thermodynamic limit values for $S(\vec{q}_{\rm peak})$ are small but finite up to 2.6~GPa.
We note that the 24$\times$12 results were excluded in the extrapolating
because they exhibited exceptional behavior, 
which we speculate to be related to the shift of the nesting vector that we find in the irreducible susceptibilities 
discussed in the previous subsection.

The extrapolated data are summarized in Fig.~\ref{fig:NqSq}(e), where we plot the thermodynamic limits of 
the peak of the charge/spin structure factors, which are defined as 
\begin{align}
C&=\lim_{N_{\rm s}\rightarrow{}\infty} N(\vec{q}_{\rm peak})/N_{\rm s} \\
M&=\lim_{N_{\rm s}\rightarrow{}\infty} S(\vec{q}_{\rm peak})/N_{\rm s}. 
\end{align}
This result indicates that charge and spin ordering are very sensitive to the pressure. 
In the low-pressure region (1 GPa $\leq P \leq$ 2 GPa), 
it is subtle whether spin- and charge-ordered states survive in the thermodynamic limit, but at least their stability is significantly suppressed.

\section{Discussion}
First, 
we discuss the relationship between the transfer integrals obtained from first-principles calculations and the distances between the
TMTTF molecules obtained from experiments. 
As shown in Figs. \ref{fig-ta} (b) and (c), increasing pressure uniformly decreases the distances 
between the molecules. 
This is consistent with the calculation in Fig.~\ref{fig:transfer_and_V}(a) showing the uniform increase in the absolute values of 
the transfer integrals. 

However, if we focus on the degree of the dimerization, a different aspect appears, i.e., the effect of the anisotropic shape of the molecular orbital. 
Experimentally, we estimated the structural dimerization $\Delta R_d$ from their intermolecular distances as in Sec.~\ref{subsec:strucure}, and its pressure dependence is shown in Fig.~\ref{fig-ta}(d).
It decreases monotonically but saturates around 3~GPa to a finite value around half of its ambient pressure value. 
In clear contrast, the degree of dimerization in the transfer integrals, $\delta t_a/\bar{t}_a$, as shown in Fig.~\ref{fig-params}(a), 
also decreases monotonically up to about 3~GPa and then saturates, but to
almost 0 with no dimerization. 
The reason for the loss of dimerization here is ascribed to the change in angle between the TMTTF molecules, as shown in Fig.~\ref{fig-ta}(e) leading to the different behavior from the simple distance effect. 
In the previous study~\cite{yoshimi2022} with comprehensive calculations for ambient pressure structures of a number of TM salts, no compound displays such a small dimerization. Even  (TMTSF)$_2$PF$_6$, in the right hand side of the unified phase diagram, which is considered to be weakly dimerized, shows $\delta t_a/t_a \sim 0.1$. This result indicates a clear difference between the effect of the chemical pressure and that of the physical pressure in TM$_2{\it X}$.

The anisotropy of the molecular orbital clearly affects the transfer integrals as well. 
In Fig. \ref{fig-ta}(c), throughout all the pressure range, the relation between interchain distances between TMTTF molecules $r_{q2}>r_b>r_{q1}$ holds and their reductions are more or less uniform. 
The interchain transfer integrals show slightly more complicated behavior 
as seen in  Figs.~\ref{fig:transfer_and_V}(a) and \ref{fig-params}(b). 
At low pressure, $|t_{q2}|< |t_{q1}|<|t_b|$ holds, which is not in the same order as in the distances above. 
While $t_b$ and $t_{q2}$ show similar pressure dependence, 
$t_{q1}$ increases rapidly than them. 
This results in the change to $|t_{q2}|<|t_b|<|t_{q1}|$ above $P=4$~GPa. 
We note that this was the reason for the variation of the Fermi surfaces and the resulting nesting vector discussed in Sec.~\ref{subsec:electronic_strucuture}. 
It is therefore clear that two elements, the distance and the in-plane tilt, are crucial for the considerations of the electronic structure. 

The pressure dependence of the bare intersite Coulomb interactions, on the other hand, almost follows that of the intermolecular distance by assuming 
that they are inversely proportional to the distances. 
This gives a justification for simplifying the intermolecular distances by taking the middle of the central C=C bonds and estimating the ratio of the Coulomb interactions (which requires integration over the MLWFs) by the inverse of these distances. 
The estimated distances in Figs.~\ref{fig-ta}(b) and \ref{fig-ta}(c) show  
 $r_{q2}>r_{b}>r_{q1} \gg r_{a1} > r_{a2}$, consistent with the intersite Coulomb repulsions with the relation  $V_{a2}>V_{a1} \gg V_{q1} > V_{b} > V_{q2}$, as shown in Figs.~\ref{fig:transfer_and_V}(b) and \ref{fig:transfer_and_V}(c), throughout the whole pressure range. 
As discussed in the literature~\cite{Yoshimi_2012,Yoshimi2012_PhysicaB}, 
 the stabilization of the pattern of CO is mostly determined by the relative strength of the intersite Coulomb interactions. 
The CO in the chain direction is first stabilized due to the large $V_{a1}$ and $V_{a2}$, 
 and then the larger value of $V_{q1}$ compared to $V_{b}$ and $V_{q2}$ leads to stabilize 
 the in-phase alignment resulting in the ferroelectric pattern of CO.
Note that, in general, the non-dimerized state is more favorable for CO stabilization, but the pressure-induced decrease in $U$ and $V$ is so large that the resulting decrease in Coulomb interactions is considered to have caused CO to disappear.

As for the screened Coulomb interactions, as emphasized in Sec.~\ref{subsec:electronic_strucuture}, they all decrease by the application of pressure. 
This is 
the screening effect by the increase of the bandwidth and is only captured by considering the electronic structure in the solid state, but cannot by taking the isolated molecule(s) and calculating the electronic state, e.g., by quantum chemistry calculations.

Finally, let us compare our mVMC results in Sec.~\ref{subsec:mVMC} with the experimental phase diagram displayed in Fig.~\ref{P-T}. 
Note that the calculation uses parameters derived from the crystal structure at room temperature but computes the zero-temperature ground states. 
Although the temperature effect is known to impact the properties of this class of materials, we can still obtain a reasonable agreement, as follows. 
The most notable point in the mVMC calculation is that the charge and spin ordering in the thermodynamic limit, stabilized at ambient pressure, is rapidly suppressed under pressure, even at $P=1$ GPa. 
The suppression is significant in the CO, while, although there are ambiguities [see the error bars in Fig.~\ref{fig:NqSq}(e)] it is more moderate in the spin ordering. 
These are consistent with the experiments, where the signature of CO in the resistivity becomes hard to track above $P=1$~GPa, while the SDW phase survives up to higher pressures. 
We note that the CO phases in (TMTTF)$_2$AsF$_6$ and (TMTTF)$_2$SbF$_6$ are shown to be unstable under the same order of magnitude of pressure~\cite{Zamborszky_2002, Yu_PRB2004}.

\section{Summary}

We investigated the pressure dependence of structural and electronic properties 
of an organic conductor (TMTTF)$_2$PF$_6$, 
by x-ray diffraction, electrical resistivity measurements, and {\it ab initio} calculations.
We obtained the full
 crystal structure data  
 up to 8~GPa by single crystal x-ray diffraction measurements at room temperature using the diamond anvil cell, 
 and from them, we evaluated the microscopic parameters of the effective model 
 using first-principles DFT calculations. 
 The derived extended Hubbard model was numerically solved with high accuracy 
 by the many-variable variational Monte-Carlo method 
 and the pressure dependence of charge and spin orderings is compared with the experimental phase diagram, especially up to 3~GPa where the detailed behavior of the resistivity was 
 investigated using the BeCu clamp-type pressure cell.

The crystal was drastically compressed up to about 3 GPa and then showed gradual shrinkage of the unit cell 
 down to 70~\% of the ambient-pressure volume at 8~GPa. 
The calculated parameters also show steep change at the low-pressure region. 
The strength of correlation, measured by the screened on-site Coulomb repulsion $U$ normalized by the mean intrachain transfer integral $\bar{t}_a$, 
 reaches the value of (TMTSF)$_2$PF$_6$ at around 1.5~GPa. 
A difference in the physical pressure compared to the chemical pressure 
 is found in the degree of dimerization: It steeply decreases by the external pressure 
 and almost disappears above 3~GPa, while it always exists for all the members of (TM)$_2${\it X} at ambient pressure. 
The ground state estimated by mVMC shows agreements with the experimental phase diagram, which both show a rapid destabilization of charge ordering by pressure, and a rather slower disappearance of the antiferromagnetic SDW state.


\section{Acknowledgements}

M.I. was supported by JSPS KAKENHI Grants No. 21740263, No. 21110517, and No. 20K05272. 
This work was supported, in part, by JSPS KAKENHI Grants No. 19H00648, No. 21340092, No. 16204022, No. 21H01041, No. 15H03681 (Y.U.), No. 22K03526 (K.Y.), No. 22H00106 (H. Mori), and No. 23H03818 (T.M.). 
The computations were performed using the facilities at the Supercomputer Center, The Institute for Solid State Physics, The University of Tokyo. 
Part of the DFT calculations were run on ACCMS, Kyoto University, with financial support from the Research Budget of Nihon University School of Medicine (T.K.). 
The authors acknowledge ISSP Supercomputer Center for providing us the data repository to store the data related to this study.

\section{DATA AVAILABILITY}
Input and output files of theoretical calculations and Crystallographic Information File (CIF) formatted files are uploaded to the ISSP Data Repository \cite{data}.
\appendix


\section{Measurement condition and crystallographic data of (TMTTF)$_2$PF$_6$}

TABLE I indicates measurement conditions of the single crystal x-ray diffraction under pressures for (TMTTF)$_2$PF$_6$ (C20H24F6PS8). 
The minimum and maximum angle (resolution); $\theta_{min}$ and $\theta_{max}$, total reflection, Independent reflection, Reflection with $I >2(I)$, $wR_2$, $wR_1 (I > 2(I))$ and GOD(goodness of fit) from 0~GPa to 8.1~GPa are listed. 
The x-ray diffraction measurements under the pressures were carried out at room temperature (293~K).
The crystal system was triclinic and the space group was $P$-1 under all measurement pressures.  
The x-ray absorption parameter $\mu$ mm$^{-1}$ (Mo $K\alpha$) at ambient pressure was 0.773 mm$^{-1}$ without DAC and  
0.771 mm$^{-1}$ with DAC, respectively.
The experimentally determined lattice parameters and angles depicted in Fig.\ref{fig-lattice} (a) and (b) are listed in TABLE II.  

\begin{figure}
    \centering
 \includegraphics[width=8.5cm]{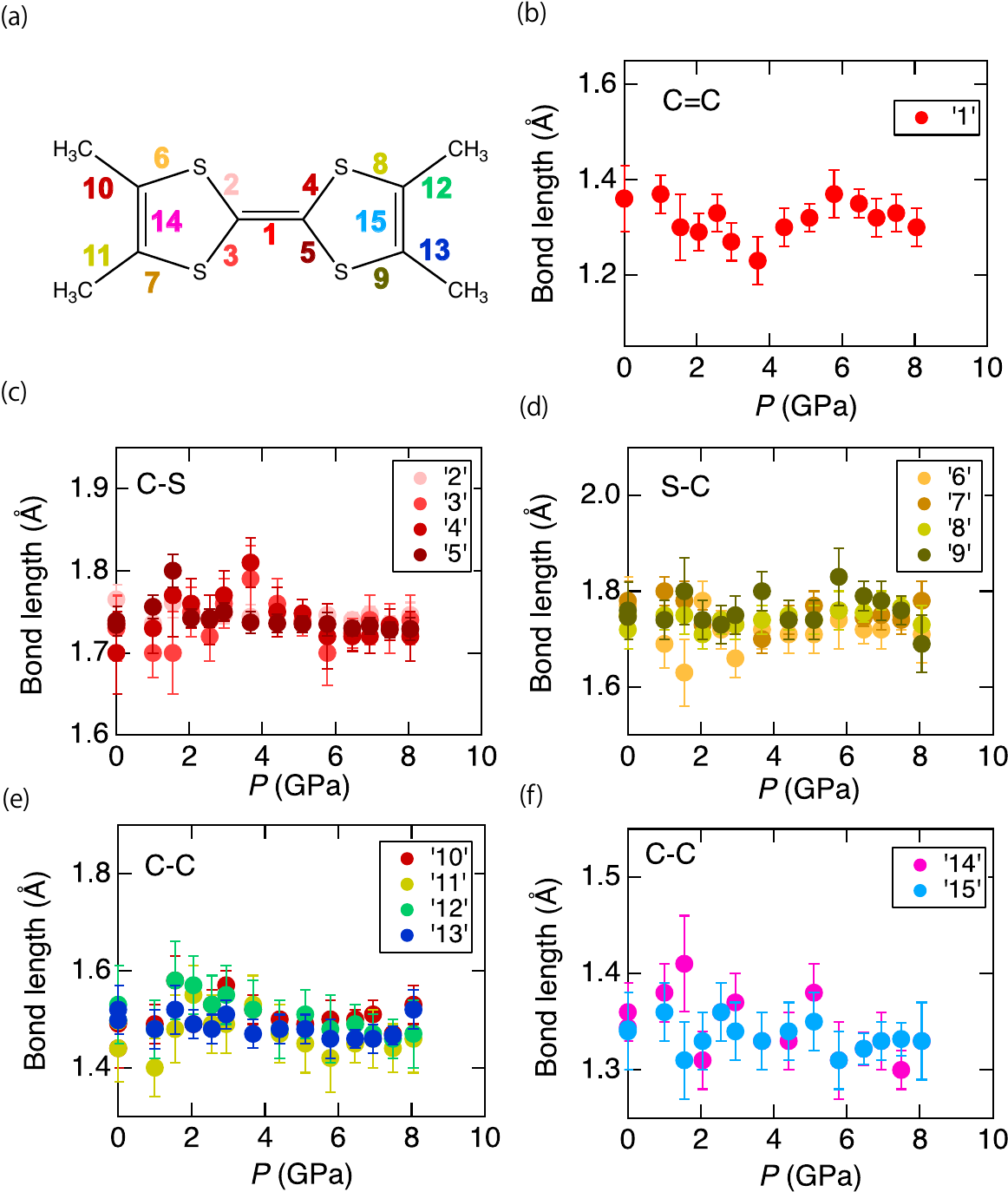}
    \caption{
   Pressure dependencies of the bond lengths C=C, C-S and C-C in TMTTF molecule. The each bond position on the TMTTF molecule is labeled with number from '1' to '15' in (a) and those variation under pressures are depicted in (b) $\sim$ (f).}
    \label{fig:L_vs_P}
\end{figure}

Pressure dependence of bond lengths C=C, C-S, and C-C in TMTTF molecules is shown in Fig.~\ref{fig:L_vs_P}. The data were obtained without constains on the bond lengths and angles during structural analysis. The data indicates that the TMTTF molecular frame hardly shrinks under pressure up to 8 GPa. The bond lengths in TMTTF at ambient pressure at room temperature without DAC are C=C: '1' 1.361(4) ~\AA, '14' 1.344(3)~\AA, '15' 1.342(3)~\AA, C-S:'2' 1.735(3)~\AA, '3' 1.731(3)~\AA, '4' 1.736(3)~\AA, '5' 1.734(3)~\AA, '6' 1.749(3)~\AA, '7' 1.748(3)~\AA, '8' 1.751(3)~\AA, '9' 1.746(3)~\AA), and C-C:'10' 1.491(4)~\AA, '11' 1.496(4)~\AA, '12' 1.498(4)~\AA, '13' 1.498(4)~\AA. 
The bond lengths in TMTTF at room temperature and 4~K for (TMTTF)$_2X$($X$=PF$_6$ and AsF$_6$) were reported in \cite{LIAUTARD} and \cite{GRANIER1988343}.

\begin{figure}
    \centering
 \includegraphics[width=7.5cm]{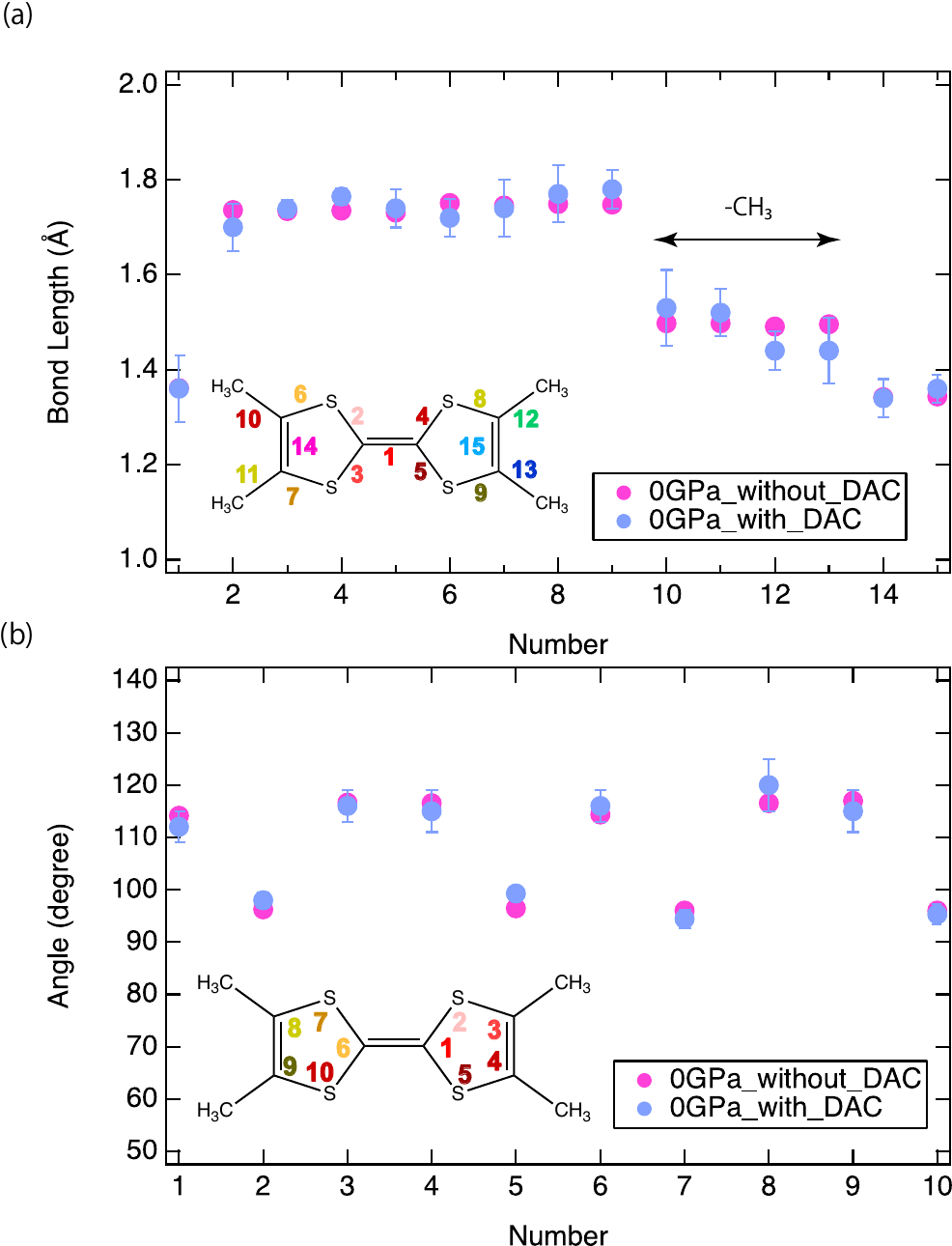}
    \caption{
   The bond lengths and angles in the TMTTF molecule for ‘0 GPa without DAC’, ‘0 GPa with DAC’. (a) The bond lengths C=C, C-S and C-C and (b) angles in TMTTF molecule. The numbers correspond to the specific positions of the bond lengths and angles. }
    \label{fig:OPT}
\end{figure}

The bond lengths and angles of the TMTTF molecule for 0 GPa with and without DAC are compared in Fig.~\ref{fig:OPT}. The error for ‘0 GPa with DAC’ is larger than ‘0 GPa without DAC’: the electron density on the S$_2$C=CS$_2$ backbone is determined with a relatively small error, and the -CH$_3$ bond at the end of the TMTTF molecule has larger error than other bonds due to the large thermal movement of -CH$_3$ (10-13). The data agree well each other.

\newpage
\begin{table*}[bt]
    \centering
     \caption{Measurement condition and crystallographic data of (TMTTF)$_2$PF$_6$ under several pressures. }
    \begin{tabular}{cccccccccc}
    \hline
     P [GPa]&$\theta_{min}$&$\theta_{max}$&Total      &Independent&Reflection     &Number of &$wR_2$&$R_1$&Goodness\\
            &   &  &Reflection&Reflection  &with $I > 2(I)$ &Parameters & &   &of fit\\
    \hline \hline
0 (without DAC)&2.993&29.888&15735&3450&2796&165&0.1109&0.0417&1.072\\
0 (with DAC)&2.987&29.126&5300&848&510&74&0.3338&0.1194&1.354\\
1.0&2.932&29.236&4221&730&516&67&0.2541&0.0913&1.114\\
1.6&3.193&25.357&3771&660&448&87&0.3285&0.1393&1.28\\
2.1&2.982&25.356&3261&637&505&74&0.1828&0.0672&1.138\\
2.6&3.003&25.373&3218&611&490&74&0.1932&0.0656&1.126\\
3.0&3.28&25.317&3283&615&503&74&0.2185&0.0669&1.134\\
3.7&3.303&25.421&2982&602&463&74&0.2146&0.0772&1.127\\
4.4&3.062&25.349&2914&587&464&74&0.188&0.0681&1.095\\
5.1&3.08&25.371&2721&574&453&75&0.1898&0.0627&1.018\\
5.8&3.356&25.262&2711&562&431&74&0.2431&0.0927&1.04\\
6.5&3.372&25.298&2837&565&470&74&0.1404&0.0505&1.094\\
7.0&3.381&25.32&2594&540&438&74&0.1682&0.0624&1.046\\
7.5&3.391&25.378&2667&554&464&74&0.1565&0.0573&1.108\\
8.1&3.398&25.45&2427&532&389&74&0.2478&0.0911&1.078\\
\hline        
    \end{tabular}
   
    \label{tab:measurement condition}
\end{table*}

\begin{table*}[bt]
     \centering
      \caption{Lattice parameters 
      of (TMTTF)$_2$PF$_6$ under different pressures: Lattice constants $a, b, c$, angle $\alpha, \beta, \gamma$ and volume $V$. 
      The ambient pressure data are listed for the conditions with and without the diamond ambient cell (DAC).} 
      
    \begin{tabular}{cccccccc}
    \hline
    $P$ (GPa)&$a$(\AA)&$b$(\AA)&$c$(\AA)&$\alpha$($^{\circ}$)&$\beta$($^{\circ}$)&$\gamma$($^{\circ}$)&$V$(\AA$^3$)\\
    \hline \hline
        0 (without DAC)&7.1513(3)&7.5756(3)&13.2083(6)&82.616(4)&84.701(4)&72.424(4)&675.42(5)\\
             0 (in DAC)&7.1586(10)&7.5828(10)&13.209(3)&82.62(2)&84.72(2)&72.566(12)&677.3(2)\\
                   1.0&6.8456(11)&7.3768(11)&12.905(3)&83.28(2)&86.54(2)&71.232(14)&612.6(2)\\
                   1.5&6.7535(10)&7.3257(10)&12.828(3)&83.51(2)&87.05(2)&70.861(13)&595.6(2)\\
                   2.1&6.6816(8)&7.2793(8)&12.754(3)&83.637(19)&87.395(19)&70.604(10)&581.50(16)\\
                   2.6&6.6178(7)&7.2368(7)&12.690(3)&83.789(17)&87.698(17)&70.409(9)&569.19(15)\\
                   3.0&6.5660(7)&7.2041(7)&12.655(3)&83.870(18)&87.937(18)&70.249(10)&560.16(15)\\
                   3.7&6.4961(9)&7.1573(9)&12.592(3)&83.93(2)&88.12(2)&70.072(12)&547.29(17)\\
                   4.4&6.4361(7)&7.1196(7)&12.542(3)&83.988(19)&88.354(18)&69.904(10)&536.73(14)\\
                   5.1&6.3799(8)&7.0834(9)&12.503(3)&84.11(2)&88.57(2)&69.787(12)&527.39(17)\\
                   5.8&6.3304(9)&7.0528(9)&12.463(3)&84.13(2)&88.65(2)&69.664(12)&518.99(17)\\
                   6.5&6.2854(7)&7.0224(7)&12.422(3)&84.167(19)&88.847(19)&69.534(10)&510.97(14)\\
                   7.0&6.2542(8)&7.0027(8)&12.393(3)&84.151(19)&88.89(2)&69.468(11)&505.58(15)\\
                  7.5&6.2246(8)&6.9819(8)&12.374(3)&84.155(19)&89.01(2)&69.392(11)&500.64(15)\\
                  8.1&6.1956(10)&6.9617(11)&12.336(4)&84.00(3)&89.09(3)&69.353(14)&495.1(2)\\    
\hline        
    \end{tabular} 
    \label{tab:Lattice parameters}
\end{table*}

\newpage

\section{Comparison with DFT-optimized structures}

\begin{figure}
    \centering
 \vspace{0.5cm} 
\includegraphics[width=9.5cm]{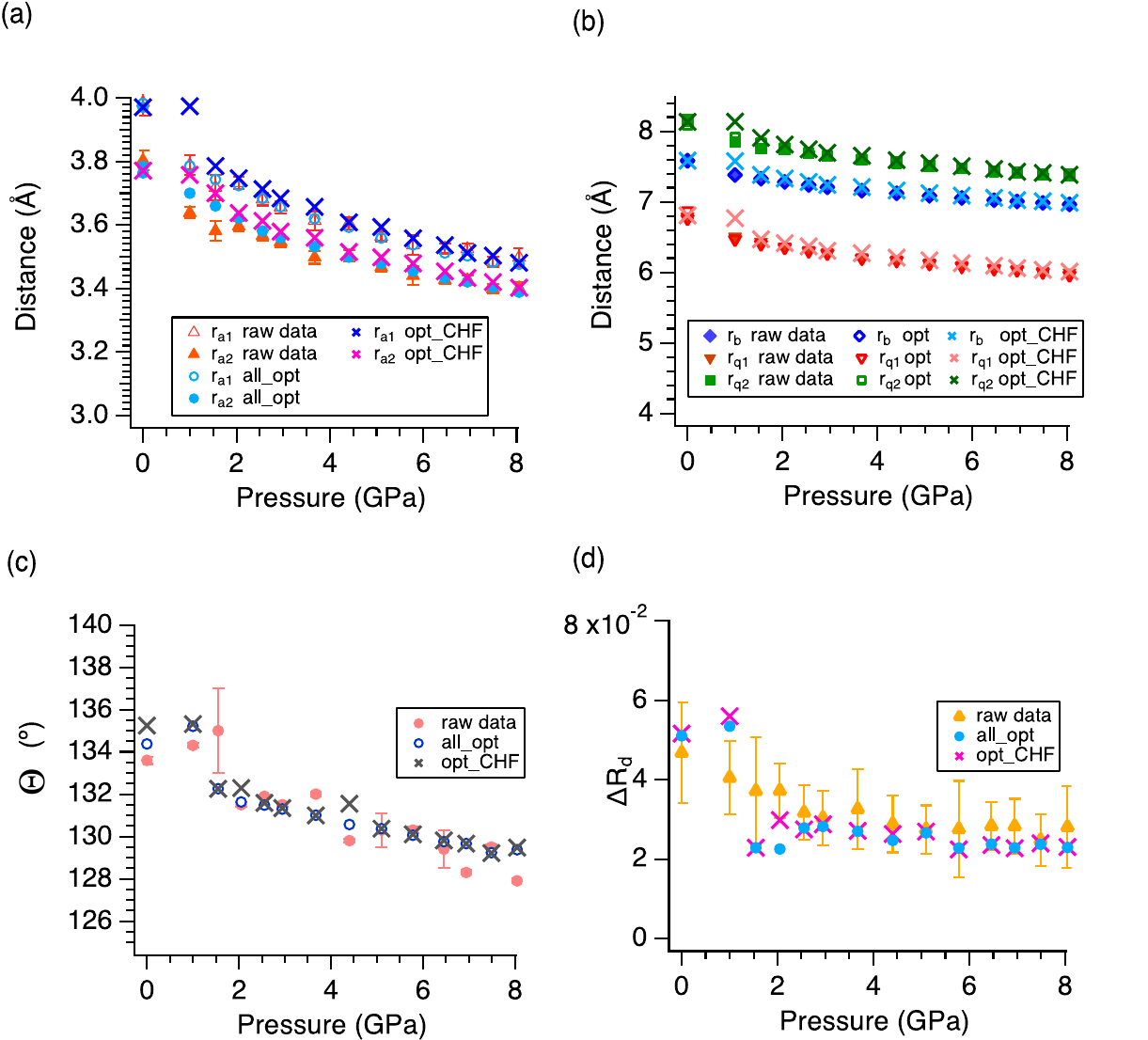}
    \caption{%
    Pressure dependencies of the distance between TMTTF molecules, 
    for the (a) intrachain and (b) interchain directions, (c) the shift angle $\Theta$, and (d) the relative dimerization $\Delta R_d$. 
    The raw data represents experimental data. The distance when all internal coordinates are relaxed using first-principles calculations is shown as all\_opt, and the results when only the atomic position of C, H and F atoms are relaxed is shown as opt\_CHF.
    The definitions of distances between molecules are the same as those 
    shown in Fig.~\ref{fig-ta}(a).}
    \label{fig:opt_I}
\end{figure}

As mentioned in Sec.~\ref{subsec:method}, to perform the theoretical analysis we first examine the relaxation of atomic potions with the first-principles DFT method, because the forces acting on H, and F atoms forming PF$_6$ and CH$_3$ were high (0.25 Ry/Bohr) within the experimentally determined structures. 
In general, C-H and C-F bond lengths determined by x-ray diffraction are shorter than the DFT-optimized structure (the optimized distances of P--F and C--H are 1.65 and 1.09~\AA, respectively)~\cite{Bhargava_PF6_2007}. 
Therefore, we performed the relaxation for C, H and F atoms with the lattice parameters at each pressure fixed by experimental data.
Figure~\ref{fig:opt_I} plots the pressure dependence on the 
intermolecular distances, together with the results when all the internal coordinates are relaxed.
By performing structural optimization, the pressure dependence of the intermolecular distance became smoother. 
Specifically, the discontinuous behavior observed at pressures below 2 GPa was eliminated~[see Figs.~\ref{fig:opt_I}(a) and \ref{fig:opt_I}(c)].
However, the relative dimerization ratio $\Delta R_d$ shown in Fig.~\ref{fig:opt_I}(d) does not match the experimental results when the optimized structure is at 2~GPa, and shows an unnatural behavior. This discrepancy is caused by the limited ability of GGA functional to accurately account for van der Waals forces, leading to a longer inter-dimer distance~\cite{vdW_Langreth05}.

\section{Chemical pressure dependence on model parameters at ambient pressure} 

\begin{figure}
    \centering
    \includegraphics[width=9cm]{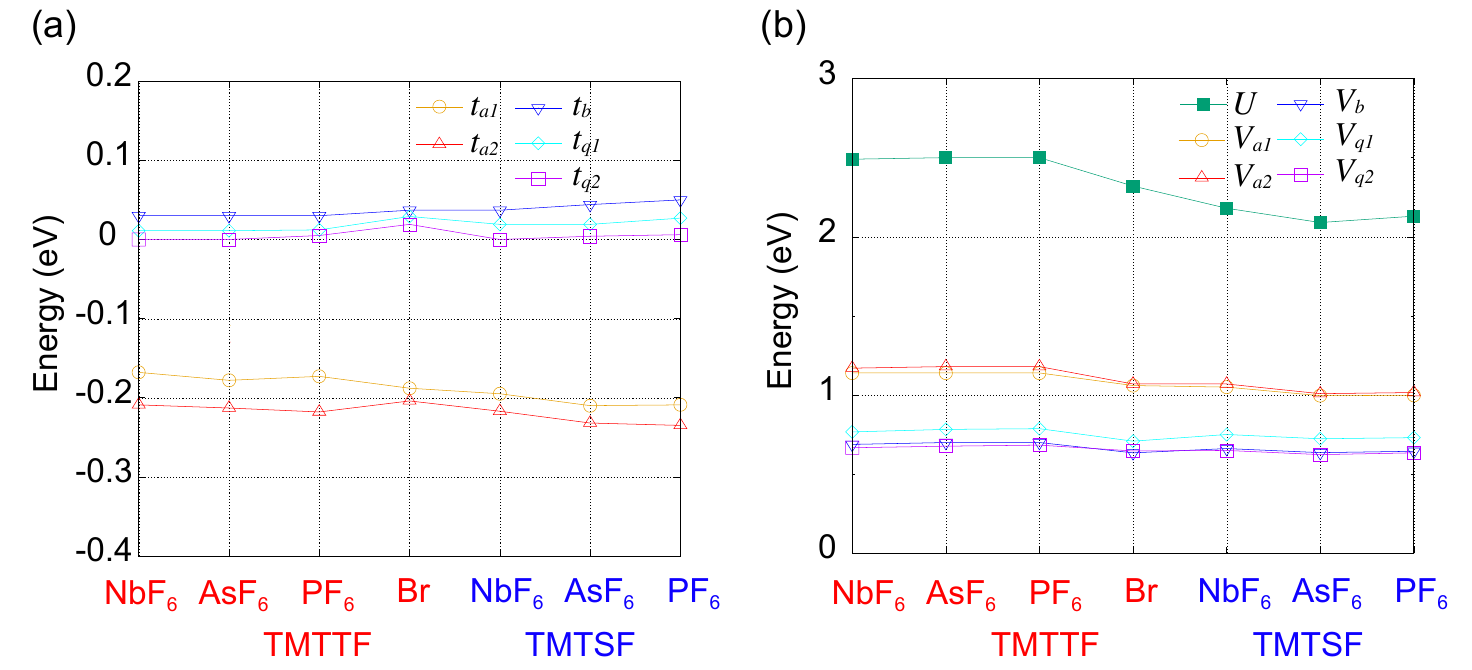}
    \caption{(a) The transfer integrals and (b) the screened Coulomb interactions for TM$_2${\it X} (TM = TMTSF, TMTTF, {\it X} = NbF$_6$, AsF$_6$, Br, and PF$_6$) at ambient pressure obtained by using the MLWFs and the DFT calculations\cite{yoshimi2022}.}
    \label{fig:model-params-chem}
\end{figure}

Figure \ref{fig:model-params-chem} shows the transfer integrals and (b) the screened Coulomb interactions for TM$_2${\it X} (TM = TMTSF, TMTTF, {\it X} = NbF$_6$, AsF$_6$, Br, and PF$_6$) at ambient pressure obtained by using the MLWFs and the DFT calculations in Ref.~\cite{yoshimi2022}.
The transfer integrals, as in the physical pressure to (TMTTF)$_2$PF$_6$ reported here, 
 basically tend to increase with increasing chemical pressure. 
However, in addition to the dimer nature remaining in the case of chemical pressure 
 as described in Sec. B, 
 the differences from the physical pressure are, in the TMTSF salts, $t_{q2}$ is almost zero and the relationship $t_b<t_{q1}$ remains unchanged.
As for the Coulomb interactions, although it tends to decrease with in both cases of chemical and  physical pressure, 
 it can be seen from Fig. \ref{fig:transfer_and_V} (b) that the physical pressure has significant effects. 
On the other hand, the relationship $V_{a2}>V_{a1} \gg V_{q1} > V_{b} > V_{q2}$ remains satisfied, similar to the physical pressure, indicating that in all cases the ferroelectric CO pattern is favored.

\bibliography{apssamp}

\end{document}